\documentclass[aps,prl,twocolumn,floatfix,superscriptaddress]{revtex4-1} 
\usepackage{graphicx}
\usepackage{amssymb,amsmath}

\usepackage{nicefrac}

\begin{document}
\title{Vacuum-assisted generation and control of atomic coherences at x-ray energies}

\author{Kilian P. Heeg}
\affiliation{Max-Planck-Institut f\"ur Kernphysik, Saupfercheckweg 1, 69117 Heidelberg, Germany}

\author{Hans-Christian Wille}
\author{Kai Schlage}
\author{Tatyana Guryeva}
\author{Daniel Schumacher}
\affiliation{Deutsches Elektronen-Synchrotron, Notkestra\ss e 85, 22607 Hamburg, Germany}

\author{Ingo Uschmann}
\affiliation{Institut f\"ur Optik und Quantenelektronik, Friedrich-Schiller-Universit\"at Jena, Max-Wien-Platz 1, 07743 Jena, Germany}
\affiliation{Helmholtz-Institut Jena, Fr\"obelstieg 3, 07743 Jena, Germany}
\author{Kai S. Schulze}
\affiliation{Helmholtz-Institut Jena, Fr\"obelstieg 3, 07743 Jena, Germany}
\author{Berit Marx}
\affiliation{Institut f\"ur Optik und Quantenelektronik, Friedrich-Schiller-Universit\"at Jena, Max-Wien-Platz 1, 07743 Jena, Germany}
\author{Tino K\"ampfer}
\affiliation{Helmholtz-Institut Jena, Fr\"obelstieg 3, 07743 Jena, Germany} 
\author{Gerhard G. Paulus}
\affiliation{Institut f\"ur Optik und Quantenelektronik, Friedrich-Schiller-Universit\"at Jena, Max-Wien-Platz 1, 07743 Jena, Germany}
\affiliation{Helmholtz-Institut Jena, Fr\"obelstieg 3, 07743 Jena, Germany}

\author{Ralf R\"ohlsberger}
\affiliation{Deutsches Elektronen-Synchrotron, Notkestra\ss e 85, 22607 Hamburg, Germany}

\author{J\"org Evers}
\affiliation{Max-Planck-Institut f\"ur Kernphysik, Saupfercheckweg 1, 69117 Heidelberg, Germany}

\date{\today}
\begin{abstract}
The control of light-matter interaction at the quantum level usually requires coherent  laser fields. But already an  exchange of virtual photons with the electromagnetic vacuum field alone can lead to quantum coherences, which subsequently suppress spontaneous emission. We demonstrate such spontaneously generated coherences (SGC) in a large ensemble of nuclei operating in the x-ray regime, resonantly coupled to a common cavity environment. The observed SGC originates from two fundamentally different mechanisms related to cooperative emission and magnetically controlled anisotropy of the cavity vacuum. This approach opens new perspectives for quantum control, quantum state engineering and simulation of quantum many-body physics in an essentially decoherence-free setting.
\end{abstract}
\maketitle

Light-matter interaction at the quantum level is ubiquitous in a multitude of modern applications and in fundamental studies on the foundations of physics alike. Even the seemingly simple process of spontaneous emission (SE) of a quantum system (e.g., an atom) turns out to be surprisingly complex \cite{Agarwal74}. It can be understood as arising from the exchange of energy between the atom and the surrounding vacuum, via the exchange of virtual photons. SE is a major obstacle in quantum engineering, as it destroys coherence. But somewhat surprisingly, the energy exchange between atom and vacuum can also create coherences between different states of the atom, if emission and re-absorption of the virtual photon occur on different transitions within the atom. In turn, these so-called spontaneously generated coherences (SGC)~\cite{Agarwal74,Ficek05,Kiffner10} enable interference between different SE channels, such that unwanted SE can be modified or even suppressed. SGC therefore are a powerful resource in quantum engineering, and numerous fascinating applications have been suggested like 
lasing without inversion \cite{Fleischhauer92b,Harris89}, 
enhancing non-linear responses \cite{Niu06}, 
quantum control of light propagation \cite{Zhou97},
quantum coherence in semiconductor-based devices \cite{Schmidt97,Faist97}, 
creation of entanglement \cite{Tang10},
stabilization of coherence in quantum computation schemes \cite{Kielpinski01,Verstraete09},
or increasing the efficiency of solar cells~\cite{Scully10,Scully11}.

\begin{figure}
\begin{center}
\centerline{\includegraphics[width=1.0\columnwidth]{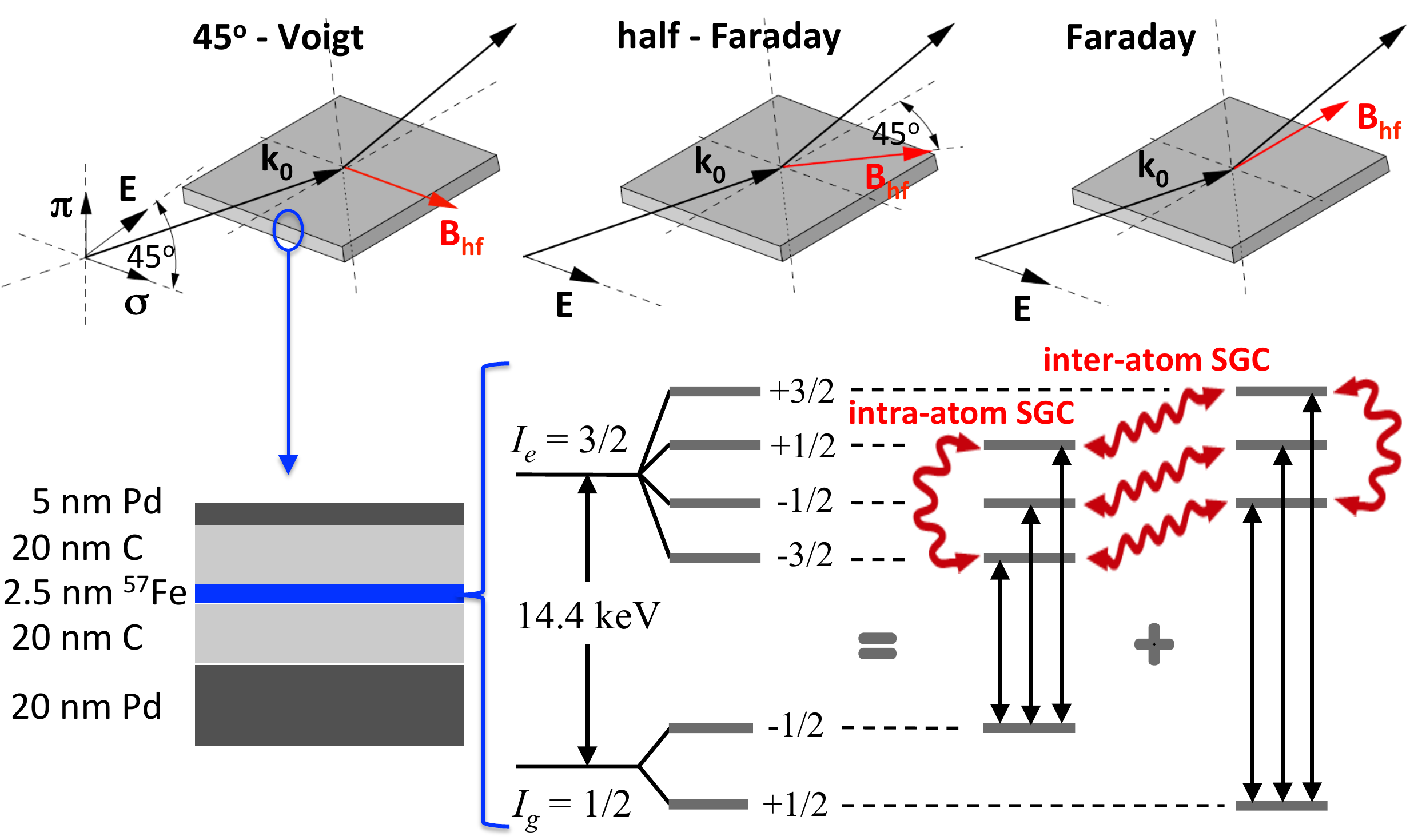}}
\caption{(Color online) Schematic representation of the sample and scattering
  geometries. The top panel shows three geometries for which the
  occurrence of SGC was investigated. Linearly polarized x-rays are
  impinging under grazing angle to evanescently couple into the
  first-order mode of the thin-film cavity (bottom left) and
  resonantly excite a thin layer of $^{57}$Fe nuclei. The
  polarization plane of the incident and the detected x-rays is
  defined by linear x-ray polarizer crystals before and behind the
  sample \cite{Toellner95,Marx11}. The magnetic hyperfine field
  $\boldsymbol{B_{hf}}$ at the position of the nuclei defines the
  quantization axis. It can be aligned and controlled via a weak
  external magnetic field. ($\boldsymbol{\sigma,\pi}$) denote the
  vectors of the linear polarization basis. The lower right panel
  shows the level scheme of the $^{57}$Fe nucleus, subject to a
  magnetic hyperfine interaction, as well as a decomposition of the
  ensemble-cavity system into an effective level scheme. Initially,
  each nucleus is in one of its two ground states, such that the
  nuclei can be divided into two distinct groups. Within each of the
  two groups, the nuclei have a single ground state and three excited
  states accessed by the probing x-ray field. SGC occur due to the
  mode anisotropy and due to interactions between different nuclei,
  and are indicated by the curly arrows between the upper
  states.}\label{fig:cavity+mechanism}
\end{center}
\end{figure}

The archetype model system for SGC is a three-level system with two upper and one common lower state (V configuration)~\cite{Agarwal74,Ficek05,Kiffner10}. If SGC between the two upper states can be established, the system can be trapped in the excited state despite its coupling to the environment. Usually, two major requirements on the structure of a quantum system, naturally not met in atoms, hinder an experimental implementation of SGC. First, the dipole moments of the transitions absorbing and emitting the virtual photon must be nonorthogonal, and second, the involved transition energies have to be near-degenerate on the level of the respective transition widths. The conditions are related to the fact that it must not be possible to know in principle which of the different decay pathways was taken \cite{Economou05}. 
It has been shown previously that observations in artificial quantum systems~\cite{Schmidt97} can be interpreted in terms of SGC~\cite{Fleischhauer92}. 
An alternative route to observing vacuum-induced coherences is to switch to $\Lambda$-type atoms in which a common excited state decays to multiple lower states. Here the requirement of non-orthogonal dipole moments can be alleviated, if an extra interaction is used to erase the knowledge to which of the different lower states the atom has decayed \cite{Aspect89,Schubert95,Weber09,Norris10}.
Alternatively, the stringent conditions of near-degenerate transitions with non-orthogonal quantum systems have recently been realized in an artificial three-level quantum system in $\Lambda$-configuration \cite{Dutt05}.
But all of these approaches have the drawback that the SE of the excited state cannot be suppressed in $\Lambda$-type setups, prohibiting full quantum control of SE, and therefore most desirable applications.

Here, we report a direct demonstration and control of V-type SGC in a cavity containing a large ensemble of $^{57}$Fe M\"ossbauer  nuclei, probed with x-rays in resonance with the nuclear transition energy of $14.4$ keV.
Embedding ensembles of $^{57}$Fe atoms in planar cavities has recently facilitated to extend quantum optical concepts into the regime of hard x-rays \cite{RR10,RR12}. 
For the observation of SGC we extend this approach by capitalizing the magnetic hyperfine splitting of the M\"ossbauer line, resulting in up to six dipole-allows transitions, see Fig.~\ref{fig:cavity+mechanism}. This not only substantially enlarges the level space available for advanced applications, but also enables control of the system via the direction and magnitude of an applied external magnetic field. The spectral response of the system is probed via the reflectivity for near-resonant x-rays impinging in grazing incidence geometry on the cavity.
In the following, we focus on three different orientations of the magnetic hyperfine field $\boldsymbol{\hat B_{hf}}$ with respect to the beam propagation direction $\boldsymbol{\hat k_0}$, the layer surface normal $\boldsymbol{\pi}$, and $\boldsymbol{ \sigma} = (\boldsymbol{\hat k_0}\times\boldsymbol{ \pi})$ as shown in Fig.~\ref{fig:cavity+mechanism}: \textit{(1) Faraday geometry:} $\boldsymbol{\hat B_{hf}} \parallel \boldsymbol{\hat k_0}$, \textit{(2) Half-Faraday geometry:} $\boldsymbol{\hat B_{hf}} \parallel \boldsymbol{\hat k_0} + \boldsymbol{ \sigma}$ and \
\textit{(3) 45$^{\circ}$-Voigt geometry:} $\boldsymbol{\hat B_{hf}} \parallel \boldsymbol{ \sigma}$. We found that in all three cases, narrow spectral dips appear in the reflectivity, which in the case of the half-Faraday and the 45$^{\circ}$-Voigt geometry lead to a vanishing signal at certain detunings. These signatures are clearly incompatible with the incoherent sum of different spectral lines, and thus point to interference that results from SGC as we will show in the following.

For the description of the observed reflected signal, a self-consistent matrix formalism has been applied (see Supplementary Material). Unfortunately, it does not provide a handle to interpret these signatures. To overcome this limitation, we have developed a full quantum optical theory for x-ray scattering from nuclei embedded in a cavity. It quantitatively agrees with previously used descriptions in the respective limits, but allowed us to clearly identify, separate and characterize all physically relevant processes contributing to the result, and furthermore provides the basis to naturally extend the modeling to non-classical light fields and non-linear light-nucleus interactions. In our ansatz, the combined system of cavity and the nuclei is modeled {\it as a single effective nucleus}, but with level structure and properties crucially modified due to cooperativity and the cavity compared to a single bare nucleus. 

The basis of our model is a consistent set of rules to choose a multi-level system, as explained in detail in the Supplementary information.  The excited level structure and the transition properties are determined by the externally controllable polarization and magnetization configuration, see Fig.~\ref{fig:cavity+mechanism}. Initially, the nuclei are incoherently distributed over the two ground states. For the determination of the linear response to the probing x-ray field, these two sub-ensembles of the nuclei can be treated separately. In the next step, we set up a master equation for the system's density matrix $\rho$ on the basis of the Hamiltonian characterizing the level structure using standard techniques~\cite{Agarwal74,Ficek05,Kiffner10}. The master equation includes individual contributions characterizing both, the single atom dynamics such as spontaneous emission, and cooperative dynamics, such as superradiance and cooperative Lamb shifts. The master equation can be written as
\begin{equation}
	\partial_t \rho = -\frac{i}{\hbar} \left [ H_\text{total},\rho \right] + \mathcal{L}^{(\text{SE})}[\rho] + \mathcal{L}^{(\text{SR})}[\rho] + \mathcal{L}^{(\text{SGC})}[\rho]\,, \nonumber
\end{equation}
where the first term on the right hand side models the coherent evolution under the total Hamiltonian $H_\text{total}$, and the second term describes spontaneous emission on individual transitions. The third term $\mathcal{L}^{(\text{SR})}[\rho]$ is another contribution to SE on individual transitions, and describes superradiant enhancement~\cite{RR10}. From our analysis we find that  a fourth term $\mathcal{L}^{(\text{SGC})}[\rho]$ has to be introduced, which leads to consistency with the experiment and the other theoretical approaches. This term involves spontaneous couplings between \textit{different} transitions, which lead to SGC.

\begin{figure}
\begin{center}
\includegraphics[width=1.0\columnwidth]{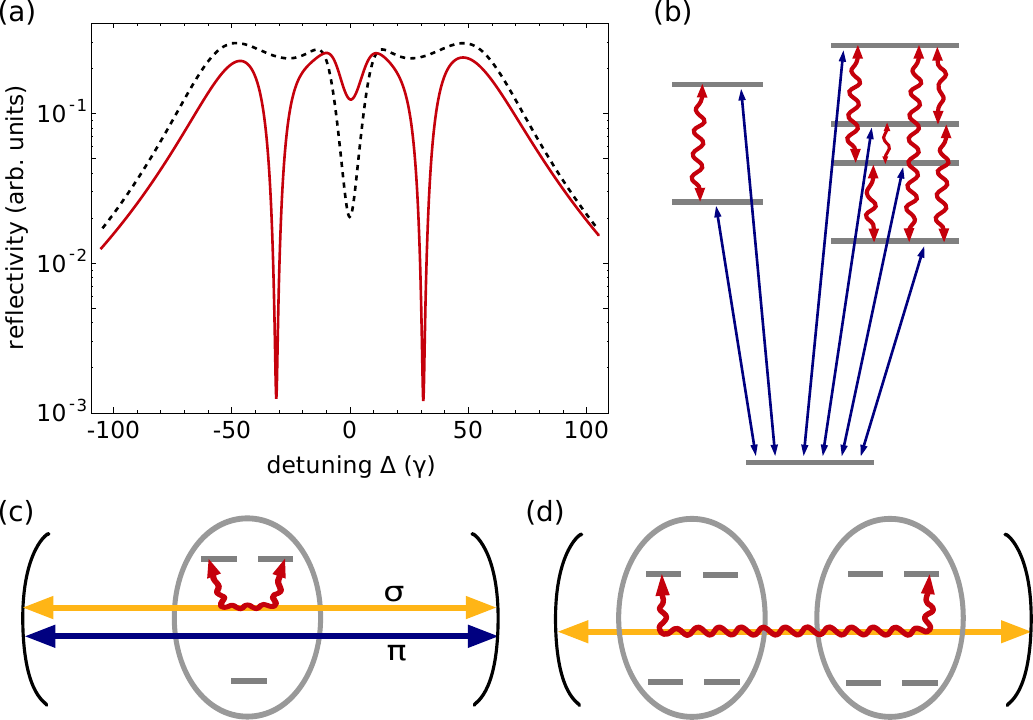}
\caption{(Color online) Origin and effect of the spontaneously generated
  coherences. {(a)} Simulated spectra for the half-Faraday geometry
  obtained from the full quantum optical model. The solid line shows
  the expected result, while the dashed line is a model calculation
  omitting the SGC contributions from the theory. The suppression of
  spontaneous emission at certain detunings due to the SGC is clearly
  visible. Parameters are $\gamma_S=27\gamma$ and $\Delta_\text{LS} =
  1\gamma$. {(b)} Effective single particle level scheme probed by the external field. The
  figure shows the case of the $45^{\circ}$-Voigt geometry. SGC
  are indicated by the curly arrows. {(c)} For certain orientations of $\boldsymbol{\hat
    B_{hf}}$ relative to $\boldsymbol{\hat k_0}$, the relevant nuclear
  transitions couple only to a single cavity polarization, giving rise
  to an anisotropic cavity vacuum and SGC between orthogonal
  transition dipoles. {(d)} In a collective effect photons are
  exchanged between different transitions in different nuclei. Probing
  the cavity as a whole results in effective SGC.} \label{fig:chi_fara45}
\end{center}
\end{figure}

The linear response of the system is obtained from the steady state solution of the master equation and yields results analytically equivalent to that from the matrix formalism. But as our quantum optical approach has the distinct advantage of separating the different contributing physical mechanisms, we can easily quantify the effect of SGC on the spectra by artificially switching them off. A detailed overview of the spectra for all considered geometries can be found in the Supplementary Material. In Fig.~\ref{fig:chi_fara45}(a) we show the effect for the \textit{Half-Faraday geometry}, where indeed the SGC lead to the narrow dips in the spectra indicative of interference.

This result raises the question, why the SGC contributions are crucial in our setting, whereas they do not contribute, e.g., for atoms in free space. Interestingly, in our setup, SGC emerge due to two fundamentally different mechanisms. 
The first contribution visualized in Fig.~\ref{fig:chi_fara45}(c) occurs on the basis of single nuclei, and arises from the fact that for certain parameter choices, the nuclei experience a spatially anisotropic photonic density of states in the cavity. To illustrate this, suppose a magnetization direction induced by $\boldsymbol{\hat B_{hf}} \parallel \boldsymbol{ \pi}$. The 2-dimensional polarization space in the cavity transverse to the propagation direction $\boldsymbol{\hat k_0}$ can be described by the orthonormal basis vectors $\boldsymbol{ \pi}$ and $\boldsymbol{ \sigma}$. In this configuration the $\Delta m = \pm 1$ transitions have dipole moments proportional to $\boldsymbol{ \sigma} \pm i \boldsymbol{ \hat k_0}$. Thus, the circularly polarized photons can only interact with the cavity mode polarized along $\boldsymbol{ \sigma}$, but not with that along $\boldsymbol{ \pi}$.
As a result, the cavity appears as having a spatially anisotropic density of states. As predicted theoretically in \cite{Zhou00,Agarwal00}, an anisotropy of this type leads to SGC. In contrast, in free space, two polarization modes would contribute, and the two (non-zero) SGC contributions of the two polarizations cancel each other.
Note that even though this effectively is a single-nucleus effect, it is assisted by cooperativity, since superradiant line broadening that is larger than the energetic splitting of the two transitions renders them indistinguishable.

The second mechanism causing SGC is a collective effect involving multiple nuclei, see Fig.~\ref{fig:chi_fara45}(d). Suppose, a photon is emitted by one nucleus with linear polarization on a $m_e=1/2 \to m_g=1/2$ transition. It can be re-absorbed in a different nucleus on the $m_g=-1/2 \to m_e=-1/2$ transition since the dipole moments are parallel. On this microscopic level, this constitutes an interaction between two different nuclei. The probe beam, however, does not resolve the dynamics of the individual nuclei, but probes the ensemble-cavity system as a whole. As a consequence, this exchange process inside the cavity appears as an effective coupling between different excited states within the level scheme of the single effective nucleus, see Fig.~\ref{fig:chi_fara45}(b). In this sense, the complicated many-body dynamics of the ensemble of nuclei mediated by the cavity acts as a ``quantum simulator''~\cite{simulator}, which mimics a single effective quantum system with properties which go beyond those of each of the individual nuclei. Here, we specifically exploit this simulation technique to induce SGC in the effective level scheme observed by the x-ray beam probing the total ensemble-cavity system.

\begin{figure}
	\begin{center}
	\includegraphics[width=1.0\columnwidth]{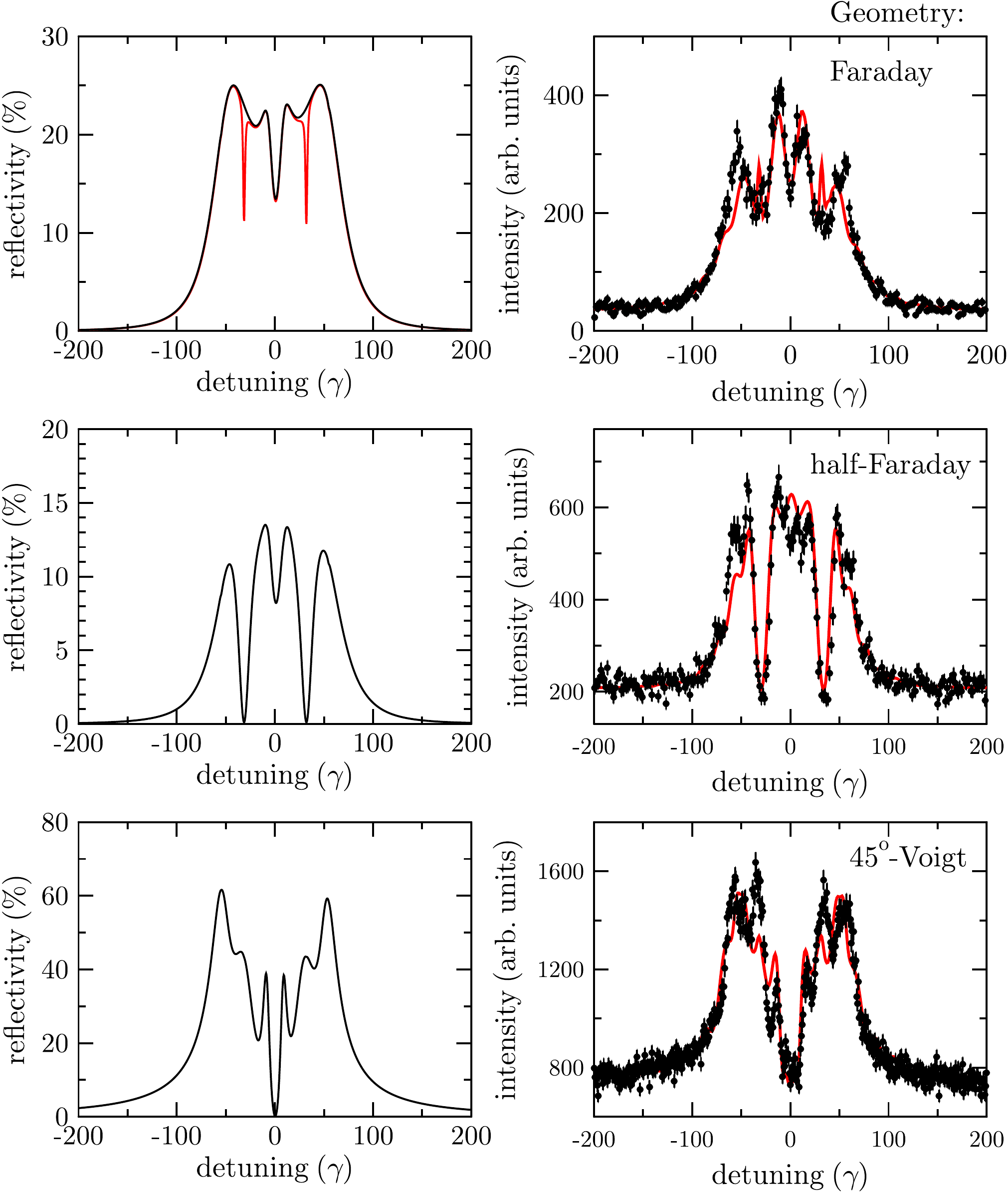}
	\caption{(Color online) Theoretical predictions and experimental results. The
          three rows show data for the three magnetization geometries
          introduced above. Left column: Theoretical predictions
          obtained from the quantum optical model. Right column:
          Experimental data and numerical simulations taking into
          account the scattering geometry, the measurement process and
          the sample parameters. The additional red curve in the top
          left panel shows the result predicted with a small angular
          deviation used to model the experimental data in the top
          right panel. The good agreement between theory and
          experiment in all cases is clearly visible. The dips in the
          reflected intensity down to the background baseline for the
          half-Faraday and the $45^{\circ}$-Voigt geometry clearly
          indicate the presence of SGC in an essentially
          decoherence-free system.}\label{fig:exp_results}
\end{center}
\end{figure}

To verify the SGC experimentally, we prepared a planar x-ray cavity consisting of a Pd(5 nm)/C(40 nm)/Pd(20 nm) layer system with the Pd layers acting as the mirrors and the C as guiding layer. A 2.5 nm thick $^{57}$Fe layer was placed in the center of the carbon layer. In order to avoid perturbing nuclear hyperfine interactions at the $^{57}$Fe/C interface, the $^{57}$Fe was sandwiched between two 0.6 nm layers of $^{56}$Fe, which in the present context has identical properties except for the resonance that we probe. In this environment the Fe layer orders ferromagnetically and the magnetic hyperfine field at the $^{57}$Fe nuclei amounts to 33.3 T.

The experiments were performed at the PETRA III synchrotron radiation source (DESY, Hamburg) employing the method of nuclear resonant scattering. This technique relies on the pulsed broadband excitation of nuclear levels followed by the time-resolved detection of the delayed photons.
To determine the energy spectrum of the cavity reflectivity, we used a method similar to that reported in  Ref.~\cite{RR10} (see Supplementary Material).
For the detection we employ two different approaches. First, an x-ray polarimetry setup was integrated into the experiment~\cite{Marx11} in order to fully take advantage of the six possible polarization-sensitive transitions resulting from the magnetic hyperfine splitting. The polarimeter consists of two Si(840) polarizer crystals in crossed setting with the sample in-between, so that it ideally only transmits photons whose polarization has been rotated ($\boldsymbol{\sigma}\rightarrow\boldsymbol{\pi}$) upon the interaction with the nuclei. This way, nonresonant background photons are suppressed by almost 10 orders of magnitude.
However, the polarimeter setup can act as an interferometer, in which the analyzer erases which-way information for different scattering channels for photons interacting with the sample. In particular, the central dip predicted for the \textit{$45^{\circ}$-Voigt geometry} caused by SGC is superimposed with an interference structure induced by the detection setup if the analyzer is used (see Supplementary Material). To clearly separate the effect of SGC, we omitted the analyzer in a second detection approach and recorded the spectrum for the \textit{$45^{\circ}$-Voigt geometry} using a high resolution monochromator for the incident light to suppress the non-resonant background. This way, all interference structures can directly be attributed to SGC.

The measured as well as the calculated spectra obtained by the quantum optical model are shown in Fig.~\ref{fig:exp_results}. Taking into account the detection technique, the numerical simulations reproduce the data very well. In particular, the deep interference minima due to SGC are clearly visible. In the case of the \textit{Faraday geometry} it turned out that the spectrum can be explained only if, quite conceivable, a slight misalignment of the internal magnetic field is assumed. As calculations indicate, this causes further minima already for small deviations from the exact Faraday geometry. Interestingly, we found that these minima also arise due to the presence of SGC. The remaining difference between the quantum optical model and the data is mainly due to time-gating effects during the measurement process.

The reduction of the reflected intensity at certain detunings can directly be traced back to the presence of non-decaying metastable excited states, formed due to the presence of spontaneously generated coherences~\cite{Zhu95}. Our measurements therefore amount to a direct observation of SGC between excited states, inducing a modified spontaneous decay. It should be noted that we observe near-perfect interference minima in the \textit{Half-Faraday} and \textit{45$^{\circ}$-Voigt geometry} in the sense that the reflected intensities drop down to the baseline. In the language of quantum optics, this indicates that the system is essentially decoherence-free over the experimental time scales, as any perturbation would inevitably lead to loss of coherence, and therefore, of a reduction of the interference leading to the SGC minimum in the spectra. The three cases in Fig.~\ref{fig:exp_results} differ only in the direction of the applied magnetization, demonstrating the external control of the system properties.

Our results not only provide an avenue to the exploitation of SGC, but also demonstrate that genuinely new systems like high-grade noise free quantum optical level schemes can be engineered in the nuclear and the optical regime alike. The capitalization of the hyperfine splitting together with a suitable choice of the polarization and the magnetization in particular enables us to realize continuously tunable and dynamically reconfigurable quantum optical level schemes in the hard x-ray regime. A single simple solid state target system thus can be manipulated dynamically and on demand to perform different tasks. The range of accessible level schemes becomes even richer if the hyperfine splitting is combined with cavities involving multiple ensembles of resonant atoms \cite{RR12}, possibly subject to individually differing magnetizations. Future setups could also involve dynamical control of the physical target structure~\cite{PhysRevA.84.043820}. 
It should be noted that our approach to realize spontaneously generated coherences is not restricted to the regime
of nuclear resonances but can be applied in the optical regime as
well, employing ensembles of resonant emitters like atoms, ions or
quantum dots that are properly placed in optical cavities.

\begin{acknowledgments}
We are grateful to R. R\"uffer for the loan of the stainless steel
analyzer foil. Moreover, we thank F. U. Dill and A. Scholl for help
with the experimental set-up. Finally we acknowledge discussions with
G. S. Agarwal and C. H. Keitel.
\end{acknowledgments}

\bibliography{SGC}{}
    
\end{document}